\documentclass[useAMS,usenatbib]{mn2e}
\usepackage{psfig}

\begin{document}
\title[Quasistars]{Quasistars: Accreting black holes inside massive envelopes}

\author[Begelman, Rossi \& Armitage] {Mitchell C. Begelman$^{1,2}$, Elena M. Rossi$^{1,3}$ \&
Philip J. Armitage$^{1,2}$ \\
$^1$JILA, University of Colorado at Boulder, 440 UCB, Boulder, CO 80309-0440 \\
$^2$Department of Astrophysical and Planetary Sciences, University of Colorado \\
$^3$Chandra Fellow \\
\tt e-mail: mitch@jila.colorado.edu; emr@jilau1.colorado.edu; pja@jilau1.colorado.edu}

\maketitle
  
\begin{abstract}
We study the structure and evolution of ``quasistars," accreting black holes embedded within massive hydrostatic gaseous envelopes. These configurations
may model the early growth of supermassive black hole seeds. The accretion rate onto the black hole adjusts so that the luminosity carried by the convective envelope equals the Eddington limit for the total mass, $M_* + M_{\rm BH} \approx M_*$. This greatly exceeds the Eddington limit for the black
hole mass alone, leading to rapid growth of the black hole. We use analytic models and numerical stellar structure calculations to study the structure and evolution of quasistars. We show that the photospheric temperature of the envelope scales as $T_{\rm ph} \propto M_{\rm BH}^{-2/5}M_*^{7/20} $, and 
decreases with time while the black hole mass increases. Once $T_{\rm ph} < 10^{4}$~K, the photospheric opacity drops precipitously and $T_{\rm ph}$ hits a limiting value, analogous to the Hayashi track for red giants and protostars, below which no hydrostatic solution for the convective envelope exists. For metal-free (Population~III) opacities this limiting temperature is approximately 4000 K. After a quasistar reaches this limiting temperature, it is rapidly dispersed by radiation pressure. We find that black hole seeds with masses between $10^3 \ M_\odot$ and $10^4 \ M_\odot$ could form via this mechanism in less than a few Myr.
\end{abstract}

 
\section{Introduction}
The formation mechanism for supermassive black holes in galactic nuclei remains unknown. Variations on most of the formation channels identified by 
\cite{begelman78} --- which include instabilities in clusters of stars or stellar remnants, and the collapse of supermassive stars or
massive discs --- are still under consideration today \citep{umemura93,freitag06,shibata02,lodato06}. A more recent idea holds that supermassive black holes result from sustained accretion onto, or mergers of, the remnants of Population~III (i.e. metal-free) stars \citep{volonteri03}, at least some of which seem likely to be massive, short-lived progenitors of stellar mass black holes \citep{carr84,bromm99,abel02,heger03,tumlinson04}.

No current observation directly constrains the different classes of model for black hole formation. However, there are clues. The existence of massive black holes in quasars at redshifts $z > 6$ strongly suggests that these black holes, at least, started forming at redshifts high enough ($z \sim 20$) to predate extensive metal enrichment by the first stars. Motivated by this, we develop in this paper a model proposed by \cite{Be06} (BVR), in which supermassive black holes form, not from Pop~III stars themselves, but rather from the evolution of a new class of Pop~III objects that might form in metal-free haloes too massive to yield individual stars. In outline, the BVR model envisages a three-stage process for black hole formation. First,
gas in metal-free haloes with a virial temperature above $T \simeq 10^4 \ {\rm K}$ flows toward the center of the potential as a result of gravitational instabilities, forming a massive, pressure-supported central object. Nuclear reactions may start, but the very high infall rate continues to compress and heat the core, precluding formation of an ordinary star. Eventually, when the core temperature attains $T \sim 5 \times 10^8 \ {\rm K}$, neutrino losses result in a catastrophic collapse of the core to a black hole. We dub the resulting structure --- comprising an initially low mass black hole
embedded within a massive, radiation-pressure supported envelope --- a {\em quasistar}. 
Initially, the black hole is much less massive than the envelope.
Over time, the black hole grows at the expense of the envelope, until finally the growing luminosity succeeds in unbinding the envelope and the seed black hole is unveiled. The key feature of this scenario is that while the black hole is embedded within the envelope, its growth is limited by the Eddington limit for the {\em whole quasistar}, rather than that appropriate for the black hole mass itself. Very rapid growth can then occur at early times, when the envelope mass greatly exceeds the black hole mass.

\begin{figure}
\psfig{figure=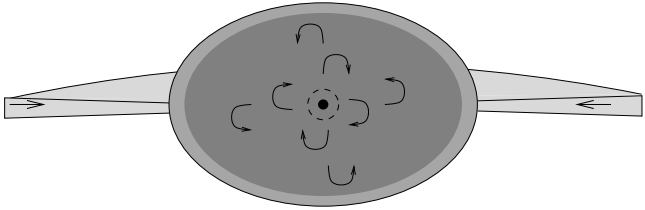,width=0.48\textwidth}
\caption{Schematic illustration of the quasistar structure that we consider in this paper. A seed black hole of mass $M_{BH}$ accretes gas from a massive,
radiation pressure-supported envelope at a rate set by the conditions outside the Bondi radius. The luminosity liberated by the accretion process is transported convectively in the inner regions of the envelope, with a transition to a radiative zone once convection becomes inefficient. In this paper we consider isolated, spherically symmetric models of quasistars, but the more physical situation would also include partial rotational support leading to flattening of the quasistar, and ongoing disc accretion at a fraction of a Solar mass per year.}
\label{fig_cartoon}
\end{figure}

The structure of quasistars, illustrated in Fig.~\ref{fig_cartoon}, resembles that of more familiar objects.
The outer regions have qualitative similarities to Thorne-\.Zytkow (1977) objects, although in 
our case the luminosity derives entirely from black hole accretion. The accretion physics is related to the ``hypercritical" regime of accretion \citep{begelman79,blondin86}, which may be observed in X-ray binaries such as SS433 \citep{begelman06}. However, these are very rough analogues. The large difference between Pop~III opacity and that of present-day, metal-enriched gas, together with the dominant role of radiation pressure and the presence of ongoing accretion in quasistars, result in a unique quasistar structure.

In this paper, we study the structure and evolution of quasistars. Using Pop~III opacities, we construct simplified analytic models which we compare with numerical integrations of the stellar structure equations. Our model quasistars are assumed to be spherically symmetric and in hydrostatic equilibrium. Real quasistars, if they exist, may not obey these strictures, and thus our results cannot be definitive. Rather, our calculations are intended to provide a first estimate of the maximum black hole mass that this mechanism can yield, together with a determination of the photospheric
temperature and luminosity of quasistars that can be used to assess their detectability with next-generation observatories such as the {\em James Webb Space Telescope}.

\section{Analytic Considerations}
We consider a black hole of mass $M_{\rm BH}$ embedded within an envelope of mass $M_* \gg M_{\rm BH}$. The luminosity of the
quasistar is generated exclusively by black hole accretion, whose rate depends upon the central conditions in the envelope. To proceed analytically, we first develop simple scaling relations  for the envelope structure and black hole accretion rate.

\subsection{Envelope Structure}
Quasistar envelopes with masses $\ga 10^3 M_\odot$ are supported primarily by radiation pressure. Since the luminosity carried by the envelope must equal the Eddington limit for the total mass, the flux carried by radiative diffusion in the envelope's interior is only a fraction $M(<r)/ M_*$ of the total, where $M(<r)$ is the mass enclosed within $r$.  Thus, quasistar envelopes are strongly convective, and their structures resemble $n=3$ ($\gamma = 4/3$) polytropes. The most accurate approach would be to model the envelope as a ``loaded polytrope'' \citep{Hu75}, with the black hole treated as a central point mass, but for $M_* \gg M_{\rm BH}$ the standard Lane-Emden solutions suffice. Defining $m_* \equiv M_* / M_\odot$, the ratio of gas to radiation pressure is uniform throughout the convective zone. We find
\begin{equation} 
 \frac{p_g}{p_r} = 7.1  m_*^{-1/2},
\end{equation}
where we have assumed a mean mass per particle $\mu \approx 0.6 m_p$, valid for fully ionized regions.  


\subsection{Energy Source}
The central regions of $n=3$ polytropes have approximately uniform densities ($\rho_c$), temperatures ($T_c$), and pressures ($p_c = aT_c^4/3$).  The boundary conditions for black hole accretion are therefore similar to those of Bondi (1952) accretion, or more specifically the generalization for optical thick flows given by \citet{flammang82}.
The adiabatic accretion rate is 
\begin{equation}
\dot M_{\rm Bondi} = \frac{4\pi}{\sqrt{2}} \frac{(G M_{\rm BH})^2 \rho_c}{c_c^3}, 
\end{equation}
where $c_c = (4p_c/ 3 \rho_c)^{1/2}$ is the adiabatic sound speed.  Bondi's solution assumes that all gravitational binding energy liberated during
accretion is advected into the hole, but this is unrealistic in the presence of even a small amount of rotation. Provided that the specific angular momentum at the Bondi radius $l_{\rm flow}$ exceeds 
the specific angular momentum $l_{\rm ms}$ of the marginally stable circular orbit, we expect the flow to be 
rotationally supported. In this case a geometrically thick accretion disk will form around the hole, within which 
angular momentum transport is required in order for accretion to occur. Although the efficiency of such a 
disk (and how the energy output is partitioned between radiation and mechanical work) is hard to calculate, 
it is reasonable to assume that it will depend primarily on the depth of the black hole potential. We write the 
luminosity as $L_{\rm BH} = \epsilon \dot M_{\rm BH} c^2$, where $\epsilon \sim O(0.1)$ is the efficiency of energy 
output\footnote{We note that $\epsilon$ {\em is} expected to vary by factors of the order of unity with the black hole spin. 
In our models, however, we are envisaging a black hole that grows in mass by several orders of magnitude. If the 
accretion flow retains a similar character during this growth, we expect that the spin of the hole will rapidly attain 
a limiting value, after which further changes will be small.} and $\dot M_{\rm BH}$ is the actual accretion rate. 
For simplicity, we take $\epsilon$ to be a constant.  
In the absence of an efficient exhaust such as a jet or evacuated funnel, this energy must be carried beyond the Bondi radius convectively, by the accreting gas itself.  Given that the convective flux density may not exceed $\sim p_c c_c$, we conclude that the accretion rate will be reduced below the Bondi value by a factor of order $\epsilon^{-1} (c_c/ c)^2 \ll 1$ (Gruzinov 1998; Blandford \& Begelman 1999; Narayan, Igumenshchev \& Abramowicz 2000; Quataert \& Gruzinov 2000). 
Unless $\epsilon \ll 0.1$ -- for example in the case where $l_{\rm flow} < l_{\rm ms}$ -- accretion is suppressed by 
a large factor, of the order of $10^4$ for the models we consider.
Combining these results we adopt the expression
\begin{equation} 
L_{\rm BH} = 4\pi G^2 \alpha M_{\rm BH}^2 \rho_c^{3/2} p_c^{-1/2}
\label{eq:lbh}
\end{equation} for the accretion luminosity. The parameter $\alpha < 1$ accounts for energy sinks within the Bondi radius: inefficient convection, presence of outflows, etc.; as well as any inefficiency of angular momentum transport. Small values of $\alpha$ imply a reduced energy supply to the quasistar.  The standard result for an $n=3$ polytrope (e.g. Hoyle \& Fowler 1963),
\begin{equation} 
\rho_c = 1.3 \times 10^{-4} m_*^{-1/2} T_6^3 \ {\rm g \ cm}^{-3}, 
\label{rhoc}
\end{equation}
 where $T_6 = T_c/ 10^6$ K, coupled with the equation of state for radiation in LTE, allows us to express $L_{\rm BH}$ in terms of the central temperature:
\begin{equation} 
L_{\rm BH} = 6.6 \times 10^{42} \alpha m_{\rm BH}^2 m_*^{-3/4}T_6^{5/2} \ {\rm erg \ s}^{-1},
\label{lbh2}
\end{equation}
where $m_{\rm BH} = M_{\rm BH}/M_\odot$. 


\subsection{Scaling Laws}
Suppose that the convective zone encompasses nearly the entire mass and radius of the envelope, allowing us to treat $M_*$ and $R_*$ as constant within the radiative layer. We will check the validity of this assumption in \S~2.4.2. We expect $L_{\rm BH}$ to be very close to the Eddington limit at the transition between the convective zone and the outer layer where radiative diffusion carries all the flux, and set $L_E = 4\pi GM_* c/ \kappa_{\rm tr}$, where $\kappa_{\rm tr}$ is the opacity at the transition radius.  As we will see below, $\kappa_{\rm tr}$ is close to the electron scattering opacity,
$\kappa_{\rm es } = 0.35$ cm$^2$ g$^{-1}$, hence $\tilde \kappa \equiv \kappa_{\rm tr}/ \kappa_{\rm es} \sim O(1)$. We therefore write
\begin{equation}
\label{lbh3}
L_{\rm BH} = 1.4 \times 10^{38} \ell_{\rm tr} \tilde \kappa^{-1} m_*  \ {\rm erg \ s}^{-1},
\end{equation}
where $\ell_{\rm tr} \approx 1$ is the Eddington factor at the transition. Equating the two expressions for $L_{\rm BH}$, we obtain
\begin{equation} 
\label{rhoc2}
T_c = 1.4 \times 10^{4} \ell_{\rm tr}^{2/5} \tilde \kappa^{-2/5} \alpha^{-2/5} m_{\rm BH}^{-4/5} m_*^{7/10} \ {\rm K } .  
\end{equation}
We are interested in systems with $m_* \sim O(100 m_{\rm BH})$, so $T_c$ typically lies in the range $10^5 - 10^6$ K and electron
scattering opacity dominates in the quasistar interior. At these temperatures, the rates of 
energetically significant thermonuclear reactions are negligible, and can be safely 
ignored\footnote{In the most massive quasistars, the central 
temperature may be high enough (a few million~K) to initiate lithium burning. This is energetically negligible, and although the presence or absence of 
lithium does affect the opacity, the effect is small for the photospheric temperatures and densities 
of interest here.}. We note that this differs from the case of envelope accretion onto 
neutron stars, where the presence of a hard surface results in high temperatures 
and potentially significant nuclear reactions within the hydrostatic region of the flow. Although we do 
not model the non-hydrostatic region of quasistars in any detail, in our case high 
temperatures {\em are} attained in the immediate vicinity of the black hole. However, even in this 
region the neglect of nuclear reactions is justified, first because black hole accretion is 
energetically much more efficient than fusion, and second because ongoing accretion limits 
the timescale over which inflowing gas is exposed to high $T$.

The polytropic relations also allow us to estimate the radius of the quasistar,
\begin{equation} 
R_* = 5.8 \times 10^{12} m_*^{1/2} T_6^{-1} \ {\rm cm},
\label{eq:r*poly}
\end{equation}
\citep{HoFo63} and therefore its photospheric (effective) temperature,
\begin{equation} 
T_{\rm ph} = 8.5 \times 10^{3}  \ell_{\rm tr}^{1/4} \tilde \kappa^{-1/4}  T_6^{1/2} \ {\rm K} . 
\label{tph}
\end{equation}
Inserting our estimate for $T_c$, we obtain
\begin{equation} 
R_* = 4.3 \times 10^{14} \ell_{\rm tr}^{-2/5} \tilde \kappa^{2/5} \alpha^{2/5} m_{\rm BH}^{4/5} m_*^{-1/5}  \ {\rm cm}  
\label{rstar2}
\end{equation}
and
\begin{equation} 
T_{\rm ph} = 1.0 \times 10^3 \ell_{\rm tr}^{9/20} \tilde \kappa^{-9/20} \alpha^{-1/5} m_{\rm BH}^{-2/5} m_*^{7/20}  \ {\rm K} .  
\label{eq:t_m_bh}
\end{equation}
Thus, quasistars should have radii of order $10^2-10^3$ AU and temperatures of a few thousand degrees.


\subsection{Radiative Layer and Photosphere}
In the outer layers of the quasistar, convection is unable to transport the total luminosity and radiative diffusion takes over as the dominant energy transport mechanism. To estimate the transition temperature between the convective and radiative zones, we note that the maximum flux that can be transported convectively is
\begin{equation}
 F_{\rm con,\,max} = \beta p c_s,
\end{equation}
where $c_s = (p/\rho)^{1/2}$ is the local (isothermal) sound speed and $\beta < 1$ is an efficiency factor. Equating this to the photospheric  flux $a c T_{\rm ph}^4/ 4$ (again assuming a narrow radiative zone), we obtain 
\begin{equation}
\label{tm}
T_{\rm tr}^4 = {3 \over 4\beta} {c\over c_s} T_{\rm ph}^4.
\end{equation} 
This result is equivalent to the condition that radiation be trapped in the convective cells as they rise.  Since $\beta < 1$ and $c/c_s \gg 1$, we deduce that $T_{\rm tr} \gg T_{\rm ph}$.  Expressing $p$ and $\rho$ in terms of $T_{\rm tr}$ (using the polytropic relations and the equation of state), we obtain
\begin{equation}
\label{tm2}
T_{\rm tr} = 31 \beta^{-2/9} m_*^{-1/18} T_{\rm ph}^{8/9} .
\end{equation} 
Since we will later show that $T_{\rm ph}$ cannot drop below a few thousand degrees, we conclude that $T_{\rm tr}$ is well above $10^4$ K. The import of this is that at the densities and temperatures likely to apply near the transition (for $m_* > 10^4$), bound-free opacity can be important but will not
elevate the Rosseland mean opacity above Thomson scattering by a large factor.
 
In order to determine the structure of the radiative layer, we need to know the behavior of the opacity. For the purpose of our analytic estimates, we adopt a simple phenomenological form for the opacity, based on the Pop III opacity tables of Mayer \& Duschl (2005, hereafter MD05).  We note that, for temperatures below $10^4$ K and densities $\la 10^{-9}$ g cm$^{-3}$, the Rosseland mean opacity depends much more sensitively on temperature than on density.  This can be seen clearly in Fig.~4 of MD05.  We therefore write
\begin{equation}
\label{modelop}
\kappa (T) = {\kappa_0 \over 1 + (T/T_0)^{-s} } .
\end{equation}
This expression does not capture the possible contribution of bound-free opacity above $10^4$~K, but it does mimic the extremely steep decline in opacity toward lower temperatures. A crude fit gives $s \approx 13$, $T_0 \approx 8000$ K, and $\kappa_0 \approx \kappa_{\rm es}$.

The advantage of using an opacity that is solely a function of temperature is that one can combine the equation of hydrostatic equilibrium with the radiation diffusion equation and integrate to obtain the total (radiation + gas) pressure as a function of temperature (see, e.g., Cox 1968, chapter
20, or any stellar structure textbook for a discussion of this technique).  Defining the Eddington factor associated with opacity $\kappa_0$ as
\begin{equation}
 \ell_0 \equiv \frac{L_{\rm BH}\kappa_0}{4 \pi G M_* c},
\end{equation}
we obtain
\begin{equation} 
\label{pint}
p (T) = {4 a \over 3 \ell_0} \left[ {T^4 - T_{\rm ph}^4 \over 4} + { T_0^s (T_{\rm ph}^{4-s} - T^{4-s})\over s-4} \right] + p_{\rm ph} ,
\end{equation}
where $p_{\rm ph}$ is the total pressure at the photosphere.  We use a standard model for the atmospheric structure, in which the radiation pressure depends on optical depth as $p_r (\tau) = (a T_{\rm ph}^4 /6) (1 + 3\tau / 2)$ and the temperature is constant between the true ($\tau = 0 $) surface and the photosphere at $\tau = 2/3$ (see, e.g., Mihalas 1978 or most stellar structure textbooks). Integrating
the equation of hydrostatic equilibrium and noting that the surface gravity can be written as $g = \kappa_0 a T_{\rm ph}^4 / (4 \ell_0)$, we have
\begin{equation} 
p_{\rm ph} = {2\over 3} {g\over \kappa(T_{\rm ph})}+ { a T_{\rm ph}^4\over 6 } = {a T_{\rm ph}^4\over 6 \ell_0} \left[ 1 + \ell_0 + \left({ T_0\over T_{\rm ph}}\right)^s   \right] . 
\label{eq:p_ph}
\end{equation}
The gas pressure is then given by 
\begin{eqnarray} 
\label{pint2}
p_ g (T) & = &  p - {a T^4 \over 3} \nonumber \\
 & = &  {4 a \over 3 \ell_0 } \left[ (1 - \ell_0) \left( {T^4 \over 4} - {T_{\rm ph}^4 \over 8}\right) \right. 
\nonumber \\ &+& \left. {T_0^s \over s-4 } \left( {s+ 4\over 8}T_{\rm ph}^{4-s} - T^{4-s}\right) \right],  
\end{eqnarray}
for temperatures $T_{\rm ph} \leq T \leq T_{\rm tr}$. 

The ratio of gas pressure to radiation pressure must be continuous across the boundary between the convective zone and the radiative layer. The matching condition is, therefore,
\begin{equation}
 \frac{3 p_g (T_{\rm tr})}{a T_{\rm tr}^4} = 7.1 m_*^{-1/2}.
\end{equation}
To simplify matters, we note
that, since $T_{\rm tr} \gg T_{\rm ph}$ and $s \gg 4$, we can neglect
the terms in $p_g (T)$ proportional to $T_{\rm ph}^4$ and $T^{4-s}$.
Solving the matching condition for $\ell_0$, we obtain
\begin{equation}
\label{ellmatch}
\ell_0 = {1 + {s+4 \over 2(s - 4)}\left({T_0\over T_{\rm tr}}\right)^s\left({T_{\rm tr}\over T_{\rm ph}}\right)^{s-4} \over 1 + 7.1 m_*^{-1/2}}  .
\end{equation}


\subsubsection{Opacity Crisis}
A necessary condition for a model to be physically realistic is that the flux not exceed the Eddington limit at the transition radius, $\ell_{\rm tr} < 1$. At the transition radius, radiation is effectively trapped, and any super-Eddington flux would be efficiently converted into bulk kinetic energy. Hence, if this condition is violated we expect the outer layers to be blown off by radiation pressure, ultimately dispersing the quasistar. For our approximate opacity, which is a monotonically increasing function of $T$, this translates to $\ell_0 < 1 + (T_0/ T_{\rm tr})^s$.  Since the convective envelope is dominated by radiation pressure, we can write this condition as
\begin{equation}
\label{ellmatch2}
 {s+4 \over 2(s - 4)}\left({T_0\over T_{\rm tr}}\right)^s\left({T_{\rm tr}\over T_{\rm ph}}\right)^{s-4} < 
 7.1 m_*^{-1/2} + \left( {T_0\over T_{\rm tr}}\right)^s.
\end{equation}
Noting that the second term on the right-hand side of this equation is negligible, we substitute for $T_{\rm tr}$ using equation~(\ref{tm2}) to obtain a lower limit on $T_{\rm ph}$,
\begin{equation}
\label{tphmin}
T_{\rm ph}^{s-4/9} > 7.7\times 10^{-8} \left( {s+ 4 \over s-4} \right) \beta^{8/9} m_*^{13/18} T_0^s.  
\end{equation}
For $s \gg 4$, this lower limit is very insensitive to the convective efficiency $\beta$ and the envelope mass $M_*$, and is roughly proportional to $T_0$. For example, adopting $s=13$ and normalizing $\beta$, $m_*$ and $T_0$ to $0.1$, $10^4$, and $8000$ K, respectively, we have
\begin{equation}
\label{tphmin2}
T_{\rm ph} > 4500 \ \beta_{-1}^{8/113} m_{*4}^{13/226} T_{0, \ 8000}^{117/113} \ {\rm K}.  
\label{eq:tmin}
\end{equation}
The corresponding lower limit on $T_{\rm tr}$ shows similar behavior:
\begin{equation}
\label{ttrmin}
T_{\rm tr} > 55,000 \ \beta_{-1}^{- 18/113} m_{*4}^{-1/226} T_{0, \ 8000}^{104/113} \ {\rm K}.  
\end{equation}

The first of these lower limits is the most important result of the analytic part of this paper. It represents a floor to the photospheric temperature of 
quasistars, analogous to the ``Hayashi track" (Hayahi \& Hoshi 1961; Hayashi 1961), which limits the temperatures of red giants and convective protostars. Our analysis differs from Hayashi's model in that our convective envelopes are radiation pressure-dominated and therefore resemble $n=3$ polytropes, rather than the $n = 3/2$ polytropes appropriate to gas pressure-dominated convection.


\subsubsection{Validity of Approximations}
The analytic model described above assumes that the growth of the black hole within 
the envelope can be described by a sequence of hydrostatic solutions. This quasi-static 
approximation is justified by the ordering of timescales. Adopting representative values 
of the mass, radius, and radiative efficiency ($m_* = 10^5$, $m_{\rm BH} = 10^3$, 
$R_* = 10^{15} \ {\rm cm}$, $\epsilon = 0.1$) the timescale on which hydrostatic 
equilibrium is established,
\begin{equation}
 t_{\rm dyn} \sim \sqrt{\frac{GM_*}{R_*}} \sim 10^8 \ {\rm s}
\end{equation}
is much shorter than the timescale on which the black hole grows
\begin{equation}
 t_{\rm grow} \equiv \frac{M_{\rm BH}}{\dot{M}_{\rm BH}} \sim 10^{13} \ {\rm s}.
\end{equation}
Similarly the timescale for attaining radiative equilibrium,
\begin{equation}
 t_{\rm rad} \sim  \frac{\tau \Delta R_*}{c} \sim 10^7 \ {\rm s},
\end{equation}
where $\tau$ is the optical depth through the radiative layer of width 
$\Delta R_*$, is enormously shorter than any of the evolutionary timescales 
of the system. 

In addition to the quasistatic assumption, we have taken the  
geometric thickness and mass of the radiative layer to be negligible compared to $R_*$ and $M_*$, respectively.  How good are these assumptions?

First, consider the mass of the radiative layer. If the layer is geometrically thin and the gravity $g$ is approximately constant across it, then we have 
\begin{equation}
\label{deltam}
{\Delta M_*\over M_*} \approx {4\pi R_*^4 a T_{\rm tr}^4 \over 3 G M_*^2} ,
\end{equation}
from the equation of hydrostatic equilibrium. Using equations (\ref{rstar2}), (\ref{eq:t_m_bh}), and (\ref{tm2}) to eliminate $R_*$, $T_{\rm tr}$, and $m_{\rm BH}$ in favor of $T_{\rm ph}$, we obtain (for $\ell_{\rm tr}= \tilde\kappa = 1$)
\begin{equation}
\label{deltam2}
{\Delta M_*\over M_*} \approx 0.2\beta_{-1}^{-8/9} m_{*4}^{-2/9} \left( {T_{\rm ph}\over 4500 \ {\rm K}} \right)^{-40/9},
\label{eq:dm_rad}
\end{equation}
where $\beta_{-1} = \beta / 0.1$.  Equation (\ref{deltam2}) indicates that the assumption of constant mass in the radiative zone is only marginally self-consistent when $T_{\rm ph}$ is close to its minimum value. The approximation improves at larger $T_{\rm ph}$.

The geometric thickness of the radiative layer is dominated by the region close to the transition temperature. The opacity is therefore close to $\kappa_0 \sim \kappa_{\rm es}$ and we may approximate
\begin{eqnarray}
\label{deltar}
 {\Delta R_*\over R_*} &\sim& {16\pi R_* a c T_{\rm tr}^4 \over 3 \kappa_{\rm es} \rho_{\rm tr}} \nonumber \\ 
 &=& 0.7 \beta_{-1}^{-2/9} m_{*4}^{-1/18} \left( {T_{\rm ph}\over 4500 \ {\rm K}} \right)^{-10/9} .
\label{eq:dr_rad}
\end{eqnarray} 
Evidently, assuming geometrical thinness in the radiative layer is an even poorer approximation than assuming constant enclosed mass. In fact, our numerical models show that even the scaling in eq.~(\ref{eq:dr_rad}) --- obtained assuming ${\Delta R_*}/{ R_*} \ll 1$ --- is not valid. The radiative layer thickness generally decreases with $T_{\rm ph}$.

\subsection{Stability}
The interior conditions of quasistars are hot enough that the opacity is dominated by electron 
scattering, yet too cool for nuclear reactions to occur. Accordingly, we do not expect any of the 
stellar instabilities that depend upon complex opacities or nuclear reaction rates to afflict 
the interior of quasistars (we note later the possibility of an instability near the surface due 
to the presence of locally super-Eddington zones). A more serious concern -- given that 
quasistars are highly radiation pressure dominated -- is dynamically instability. In a 
non-rotating model, relativistic effects raise the critical $\gamma$, below which instability 
occurs, to
\begin{equation}
 \gamma_{\rm crit} = \frac{4}{3} + \delta \gamma_{\rm crit}
\end{equation}
where $\delta \gamma_{\rm crit} \sim GM / Rc^2$ (e.g. Shapiro \& Teukolsky 1983). This must be compared to the 
actual $\gamma$ in the radiation dominated interior of the quasistar. For $p_g / p \ll 1$, the 
first adiabatic exponent is given by
\begin{equation}
 \Gamma_1 \simeq \frac{4}{3} + \delta \gamma = \frac{4}{3} + \frac{1}{6} \frac{p_g}{p}. 
\end{equation}
If the pressure-weighted integral over the star of $\delta \gamma < \delta \gamma_{\rm crit}$, instability 
is possible.

The region of the quasistar interior to the Bondi radius is not in hydrostatic equilibrium, 
so it is meaningless to apply the above condition there. For an estimate, we  
evaluate the above expressions at the Bondi radius using the expressions for the interior 
structure of the quasistar. We find that the change in the stability boundary is 
\begin{equation} 
 \delta \gamma_{\rm crit} = \left( \frac{c_s}{c} \right)^2
 = 4 \times 10^{-10} m_*^{6/5} m_{\rm BH}^{-4/5} \alpha^{-2/5}
\end{equation}
while the actual $\gamma$ exceeds $4/3$ by an amount,
\begin{equation}
 \delta \gamma \simeq 1.2 m_*^{-1/2}.
\end{equation}
We conclude that instability is possible if $m_*$ is large enough. For 
$\alpha = 0.01$ and $m_{BH} = 10^3$, for example, instability is possible 
for $m_* > 3 \times 10^6$.

We caution that this analysis is far too simple to describe the actual situation. 
Globally, a robust expectation is that quasistars will be rapidly rotating as a consequence 
of their formation from rotationally supported gas (BVR). Unfortunately,  this does 
not help us in estimating the rotation rate at the Bondi radius, as the latter depends 
upon the very uncertain role of convection in redistributing angular momentum. 
However, any significant rotation would help stabilize the structure against dynamical instability.


\section{Numerical Models}
Although our analytic models ought to yield a reasonably accurate picture of the generic features of quasistar envelopes, they are only marginally self-consistent for photospheric temperatures  approaching the floor value. Moreover, real Pop~III opacities, which are functions of density as well as of temperature, have considerably more structure than the analytic fit employed in the analytic models, and require a numerical treatment.


\subsection{Equations}
\label{sec:num_eqs}
We model quasistars as static, spherically symmetric objects in thermal equilibrium. The equation of hydrostatic equilibrium is
\begin{equation}
\frac{dp}{dr} = - \frac{G\,M(r)}{r^2} \rho,
\label{eq:hye}
\end{equation}
where the mass enclosed within radius $r$ is given by
\begin{equation}
\frac{dM}{dr} = 4 \pi r^2 \rho. 
\label{eq:dmdr}
\end{equation}
At the Bondi radius, which for our numerical integrations we define as 
\begin{equation}
 R_{\rm Bondi} \equiv \frac{G M_{\rm BH} {\rho_c}}{2p_c}, 
\end{equation}
we set $M(R_{\rm Bondi})=M_{\rm BH}$. The gas mass within the Bondi radius is 
negligibly small compared to the black hole mass, so ignoring the non-hydrostatic 
region in our structure calculations should be an excellent approximation. 
The equation of state is a mix of gas and radiation pressure, 
\begin{equation}
p= p_g + p_r = \frac{\rho k T}{\mu} + \frac{1}{3} a T^{4},
\label{eq:eqstate}
\end{equation}
where $k$ is the Boltzmann constant and $\mu$ the mean molecular weight. We set $\mu = 0.6 m_p$, appropriate for a fully ionized gas, and ignore the variation of $\mu$ due to partial ionization near the quasistar photosphere.  This is a very good approximation, since the neutral hydrogen fraction generally does not exceed a few per cent.  

The temperature gradient is determined by the relative amount of energy carried by convection $(F_{\rm con})$ and radiation $(F_{\rm rad})$ at each radius.  Since quasistars are powered solely by black hole accretion, the luminosity $L_{\rm BH} = 4 \pi r^2 (F_{\rm con}+F_{\rm rad})$ is constant 
with radius. To determine the boundary of the convective zone, we use the Schwarzschild criterion for convective stability\footnote{Note that, in using the Schwarzschild rather than the Ledoux criterion for convective  stability, we again ignore effects due to changes in the mean molecular weight with radius. As explained above, we believe that this is a good approximation.},
\begin{equation}
\frac{dT_{\rm rad}}{dr} - \frac{dT_{\rm ad}}{dr} > 0,
\label{eq:schw}
\end{equation} 
where the radiative gradient $dT_{\rm rad}/{dr}$ is the gradient of temperature if all the energy is transported by radiative diffusion, 
\begin{equation}
 \frac{dT_{\rm rad}}{dr} = - \frac{3}{4 a c}\,\frac{L_{\rm BH}}{4 \pi r^2}\, \frac{\rho \kappa}{T^3},
 \label{eq:dtdr_rad}
\end{equation}
and the adiabatic gradient, 
\begin{equation}
\frac{dT_{\rm ad}}{dr} = \left( \frac{\Gamma_{2} - 1}{\Gamma_{2}} \right) \frac{T}{p} \frac{dp}{dr},
\end{equation}
describes the temperature variation in a convective element as it moves adiabatically (no radiative losses) through the surrounding layers. The adiabatic exponent $\Gamma_2$ is given by 
\begin{equation} 
 \Gamma_{2} =\frac{32 - 24\beta -3\beta^2}{24 -18\beta - 3\beta^2},
\end{equation}
where $\beta = {p_g}/{p}$ is the fraction of the total pressure contributed by gas pressure. We then set the {\it actual} temperature gradient to be, 
\begin{equation}
 \frac{dT}{dr} = \frac{dT_{\rm rad}}{dr} - \frac{\left[\min(-\frac{dT_{\rm rad}}{dr},- 
 \frac{dT_{\rm ad}}{dr}) + \frac{dT_{\rm rad}} {dr}\right]}{(1+x^{10})},
\label{eq:dtdr}
\end{equation} 
where $x = F / F_{\rm con,\ max}$. Typically, we find that the pressure scale height (which in a mixing length 
theory for convection is comparable to the distance convective elements travel) is a small fraction of the 
radius across the outer $\approx$90\% of the quasistar by radius. In the inner regions, where the pressure 
profile is quite flat, $x$ is very small, and hence it is reasonable to assume that convection (even if it 
behaves differently from a mixing length theory) also succeeds in establishing the adiabatic gradient there.
As in the analytic models, we note that the maximum flux that convection can transport is limited by the condition that convective motions cannot be supersonic. Accordingly, we set
\begin{equation} 
 F_{\rm con,\ max} = \beta c_s p_r 
\end{equation}
with the efficiency parameter $\beta = 0.1$.

Equation~(\ref{eq:dtdr}) describes the limiting behaviors of the temperature gradient in the different regions of a quasistar. The adiabatic gradient applies if the region is dynamically unstable and the flux is less than the maximum convective flux; otherwise, the radiative gradient applies. These limiting cases accurately model the regions of high convective efficiency, in the denser and more opaque interior layers, and of high radiative efficiency, in the generally stable outer layer. In the transition region, however, our use of a flux limiter (the $1+x^{10}$ term in eq.~[\ref{eq:dtdr}]) is only an approximation. A more accurate treatment would make use of mixing length theory. However, we have checked that different descriptions of the transition region do not affect appreciably the overall quasistar structure.


\subsection{Integration Method} 
\begin{figure}
\psfig{file=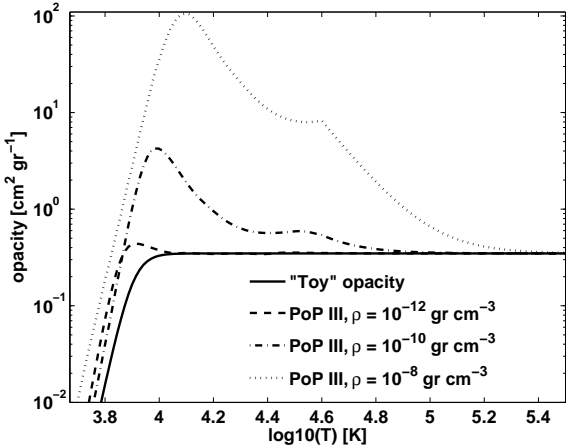,width=0.48\textwidth}
\caption[]{Opacity versus temperature. The Pop~III opacity from MD05 is plotted as a function of temperature for three different values of matter density as shown in the legend.  The range $10^{-13} \ {\rm g \ cm^{-3}} \le \rho \le 10^{-8} \ {\rm g \ cm^{-3}}$ spans the densities to be found in the outermost layers of quasistars. The solid line shows the analytic opacity given by equation~(\ref{modelop}). The figure shows how the Pop~III opacity
increases over the analytic fit as the density increases. The effect is especially evident around $T=10^{4}$~K, where there is a bound-free peak due to hydrogen ionization.}
\label{fig:opa_comp}
\end{figure}

Equations~(\ref{eq:hye}), (\ref{eq:dmdr}), (\ref{eq:eqstate}), and (\ref{eq:dtdr}), together with the definitions of the radiative and adiabatic temperature gradients, form a closed set of equations for the 4 unknowns ($p$, $T$, $\rho$ and mass), given a luminosity $L_{\rm BH}$ and stellar
radius $R_*$. To solve this system, we fix the photospheric temperature $T_{\rm ph}$, the black holes mass $M_{\rm BH}$, and the parameter $\alpha$,  and guess the photospheric radius $R_*$ and the quasistar mass $M_*$. The luminosity is then $L_{\rm eff} = 4 \pi R_*^2 \sigma T_{\rm ph}^4 $, where $\sigma = ac/4$. We determine the photospheric pressure and density using the first equality in eq.~(\ref{eq:p_ph}), together with the equation of state (eq.~[\ref{eq:eqstate}]). We then integrate inward from the known photospheric conditions, adjusting our initial guesses until we match the two conditions
\begin{eqnarray} 
 L_{\rm eff} &=& L_{\rm BH}(10 \times R_{\rm Bondi}) \\
 M_*(R_{\rm Bondi}) &=& 0.
\end{eqnarray} 
We calculate $L_{\rm BH}$ (eq.~[\ref{eq:lbh}]) with $\rho$ and $p$ evaluated at $10 \times R_{\rm Bondi}$, well beyond the inner region affected by the black hole gravity. For our numerical examples $\alpha$ is assumed to equal $0.1$, if not otherwise stated. We discuss the dependence of our results on  $\alpha$ in \S~\ref{sec:alpha}.  We explore two models for the opacity: the ``toy" opacity given by eq.~(\ref{modelop}), which we used previously in the analytic models, and the full numerical opacity tabulated by MD05, which we refer to as ``Pop~III opacity".  Fig.~\ref{fig:opa_comp} shows a comparison of these opacities for different densities of interest in the outermost radiative layers.

\subsection{``Toy" Opacity}
\label{sec:toyo}
As a test, both of our analytic models and of our numerical scheme, we first compute structure models using the ``toy" opacity (eq.~\ref{modelop}). The toy opacity ignores density dependence, and in particular the contribution from bound-free and free-free absorption at $T> 8 \times 10^3$~K. As can be seen in Fig.~\ref{fig:opa_comp}, this is a reasonable approximation at low density, but is very poor at higher density.

\begin{figure}
\psfig{file=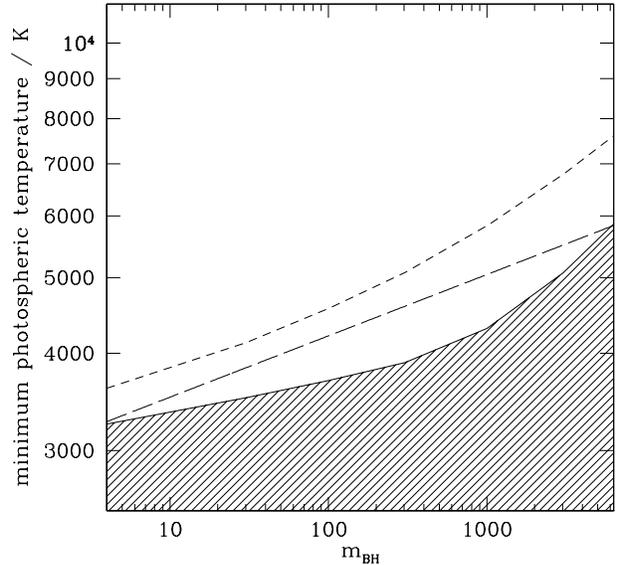,width=0.48\textwidth}
\caption[]{Minimum photospheric temperature $T_{\rm min}$ versus black hole mass in unit of solar masses $m_{\rm BH}$, for $\alpha =0.1$.  Numerical models with the "toy" opacity (short-dashed line) are compared to results computed using the Pop~III opacity (solid line). The analytic estimate, obtained by combining equations (\ref{eq:tmin}) and (\ref{eq:t_m_bh}), is shown as the long-dashed line.  Static quasistar models do not exist in the lower shaded region.}
\label{fig:tmin_vs_mbh}
\end{figure}

As expected, we find that the analytic scalings are most reliable for high photospheric temperatures, and less reliable close to the temperature floor.  We confirm that for a given ($\alpha, M_{\rm BH}$) there is indeed a minimum photospheric temperature, $T_{\rm min}$, above which the luminosity at the transition radius is sub-Eddington. Below $T_{\rm min}$, the whole radiative layer experiences a radiative force greater than the gravitational force and the quasistar is bound to evaporate. In Fig.~\ref{fig:tmin_vs_mbh}, the short-dashed line shows $T_{\rm min}$ as a function of $M_{\rm BH}$ for numerical models computed using the toy opacity. The analytic estimate, obtained by combining equations (\ref{eq:tmin}) and (\ref{eq:t_m_bh}), is shown
as the long-dashed line. Numerically, we obtain a slightly higher $T_{\rm min}$. The discrepancy is higher for higher black hole masses.  We note that for models computed using the toy opacity, the critical temperature below which the {\em entire} radiative layer becomes super-Eddington (our definition of $T_{\rm min}$) is close to the temperature at which {\em any point} in the radiative layer becomes super-Eddington. This simple behavior, which occurs because the toy opacity is constant in most of the radiative layer before dropping off steeply and monotonically near the photosphere, does not carry over to models computed using full Pop~III opacity.

\subsection{Pop III Quasistars}
\label{sec:PoPII}
To extend these results, we compute numerical quasistar structures with realistic opacities. We assume that quasistars have a primordial (metal-free) nuclear composition, and use the opacity table calculated by MD05. This table covers the density range $10^{-16}  < \rho\; [{\rm g\, cm^{-3}}] <
10^{-2}$ for temperatures $63 < T\;  [{\rm K}] < 4 \times 10^{4}$. To compute the interior structure of the quasistar --- where temperatures attain values much higher than $4 \times 10^4$~K --- we analytically extend the opacity table assuming that the excess opacity above the electron scattering value has a Kramers form, i.e., that,
\begin{equation} 
 \kappa(\rho,T) = C(\rho) \rho T^{-3.5} + \kappa_{\rm es}.
\end{equation} 
We fix the function $C(\rho)$ to match smoothly onto the tabulated opacity at the highest temperature point. As expected, Thomson scattering is the only significant source of opacity deep within the interior. 

As is evident from our Fig.~\ref{fig:opa_comp}, or from Fig.~4a of MD05, the opacity deviates substantially from our toy model around $T \simeq 10^4$ K for $\rho > 10^{-12} \ {\rm g\, cm^{-3}}$. The prominent enhancement of the opacity with increasing density is due to neutral hydrogen absorption, ${\rm H} + h\nu \rightarrow {\rm H}^+ + {\rm e}^{-}$. 

The differences between the toy and full Pop~III opacities are significant only at low temperatures. In the deep interior of the quasistar the opacity is constant, as it was for the analytic opacity, and we again find a convective zone that is described by an $n=3$ polytrope for which  ${p_g}/{p_r} \simeq 7.1 m_*^{-0.5}$. 

In the outer region the flux is transported by radiative diffusion.  The radiative layer covers between $\sim$ 50 per cent and $\sim$ 10 per cent of the quasistar by radius, becoming thinner as the photospheric temperature decreases toward $T_{\rm min}$ (for a fixed $m_{\rm BH}$) or as the black hole becomes more massive (for a fixed $T_{\rm ph}$).  In contrast, the percentage of mass in the outer radiative layer increases as the limiting temperature is approached. For $m_{\rm BH} = 300$ it goes from $\sim$ 1 per cent at $T_{\rm ph}=10^{4}$~K to $\sim 18$ per cent around T$_{\rm min}$, while for $m_{\rm BH} = 7 \times 10^{3}$ it shows a similar behavior but never exceeds 2 per cent.  

\begin{figure}
\psfig{file=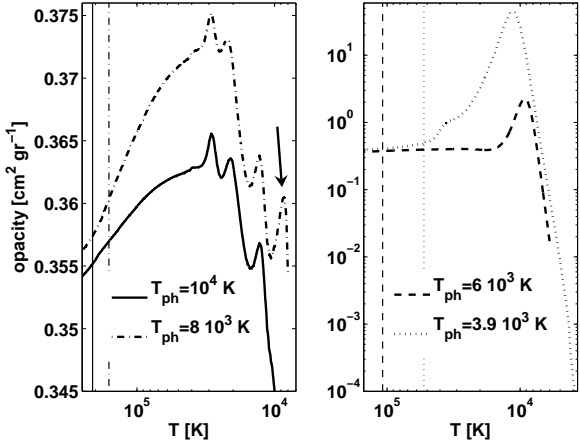,width=0.48\textwidth}
\caption[]{The opacity as a function of temperature in the outer layers of four quasistar models with M$_{\rm BH} = 300 M_{\odot}$ and $\alpha =0.1$, computed usinf Pop~III opacities. The photospheric temperatures $T_{\rm ph}$ of the models are shown in the legends. On the right-hand panel we adopt a
logarithmic scale for clarity, while the plot on the left-hand panel uses a linear scale. The temperature at the transition between the convective and radiative zones is denoted by the corresponding vertical lines. Note the rise in the bound-free peak at around $T \simeq 10^4$~K (marked on the left panel by an arrow), which becomes dominant as the photospheric temperature drops and the photospheric density increases (see also Figs.~\ref{fig:tr} and \ref{fig:rhor}). Note also the steep drop in opacity at the photospheres.}
\label{fig:kappat}
\end{figure}
\begin{figure}
\psfig{file=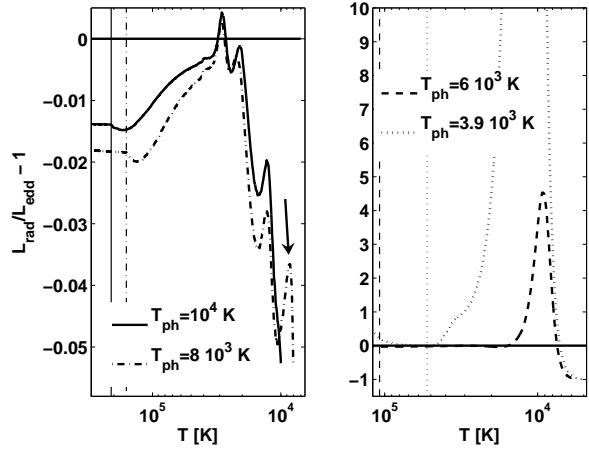,width=0.48\textwidth}
\caption[]{$L_{\rm rad}/L_{\rm edd}-1$ for the four models shown in Fig.~\ref{fig:kappat}.  The solid horizontal line marks the border between the super-Eddington (above) and the sub-Eddington (below) regions.  Note that for $T_{\rm ph} \sim 10^{4}$~K the super-Eddington zone is very narrow and confined around $T \sim 3 \times 10^{4}$~K.  For $T_{\rm ph}< 10^{4}$~K a peak appears around $T \sim 10^{4}$~K (marked by an arrow in the left-hand panel) and expands, eventually encompassing the whole radiative layer (dotted-line, right-hand panel).}
\label{fig:lradleddt}
\end{figure}

The dramatic differences between structures computed using the real opacity, and those based on the toy opacity, are almost exclusively confined to the radiative layer. In Fig.~\ref{fig:kappat} and Fig.~\ref{fig:lradleddt}, we show examples of the behavior of the opacity and the local Eddington ratio, $L_{\rm rad}/L_{\rm edd}$ with 
$L_{\rm edd} = 4 \pi G M(r)\,c/ \kappa(r)$, for models computed with four different photospheric temperatures and constant black hole mass.  The plots focus on the radiative zone. For clarity we plot temperature on the $x$-axis, since it spans more than an order of magnitude while the radius increases only by tenths of a per cent. As $T_{\rm ph}$ decreases, the opacity increases everywhere and the growth of the bound-free peak around $T = 10^4$~K becomes particularly
prominent. Correspondingly, a peak in the ${L_{\rm rad}}/{L_{\rm edd}}$ ratio forms. This peak first becomes super-Eddington for $T \simeq 10^4$~K, and steadily grows and expands as the photospheric temperature drops until the entire radiative zone is super-Eddington at $T_{\rm ph}=T_{\rm min}\simeq 4 \times 10^3$~K.  Note also that for $T_{\rm ph} \sim 10^4$~K a narrow super-Eddington region exists around $T\sim 3 \times 10^{4}$~K.  This is also true for higher photospheric temperatures. We comment on this point at the end of this section.

Fig.~\ref{fig:kappat} clearly shows the steep drop in opacity at the photosphere, and how it becomes more vertiginous as the quasistar cools down to $T_{\rm min}$.  This feature ensures (as for the ``toy'' opacity) the presence of a minimum temperature, below which no hydrostatic solutions can be found
(Fig.~\ref{fig:tmin_vs_mbh}).  As $T_{\rm ph}$ decreases, the boundary condition at the photosphere (eq.~[\ref{eq:p_ph}]) implies that the photospheric pressure passes through a minimum  and then begins to increase.  The photospheric radius {\it and} the transition radius $R_{\rm tr}$ then stop increasing and start sinking rapidly inward in order to maintain hydrostatic balance. Eventually no hydrostatic solution is possible. 

\begin{figure}
\psfig{file=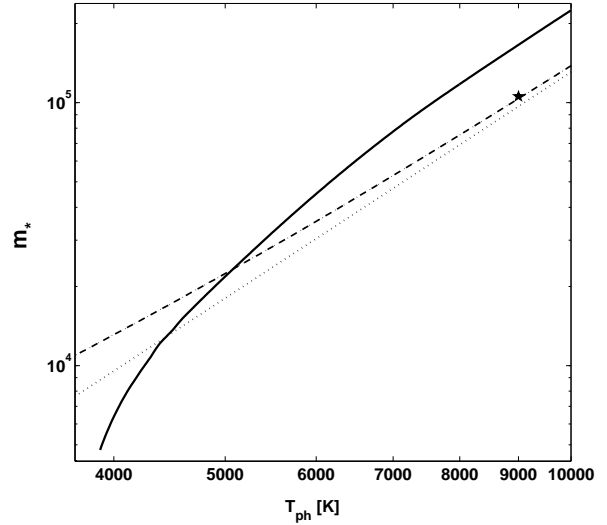,width=0.48\textwidth}
\caption[]{The quasistar mass is plotted as a function of the photospheric temperature (solid line) for models with m$_{\rm BH}$ = 300 M$_{\odot}$ and $\alpha =0.1$, computed with full Pop~III opacities.  The dotted line shows the analytic scaling $m_* \propto T_{\rm ph}^{20/7}$, obtained assuming that $\ell_{\rm tr}= \tilde \kappa=1$. The dashed line is also derived from the analytic relation  given in eq.~(\ref{eq:t_m_bh}), but with numerical values for $\ell_{\rm tr}$ and $\tilde \kappa$ used instead of assuming that these factors are exactly unity. The star shows the mass of an 
(unphysical) numerical model computed with no radiative zone. This closely matches the analytic prediction, and indicates that the offset at high temperatures is due to the presence of a radiative zone.}
\label{fig:m_vs_tph}
\end{figure}

Despite the complicated behavior of the opacity, the analytic power-law scalings between quasistar mass, photospheric temperature, and black hole mass (eq.~[\ref{eq:t_m_bh}]) remain reliable at sufficiently high temperatures, although they suffer an offset in normalization correlated with the radial thickness of the radiative layer.  To illustrate this, Fig.~\ref{fig:m_vs_tph} shows the numerically computed $m_*$ vs $T_{\rm ph}$ for models with M$_{\rm BH}= 300 \ M_\odot$.  The analytic scaling,
\begin{equation} 
 m_* \propto T_{\rm ph}^{20/7}
\end{equation} 
holds well down to $T_{\rm ph} \sim 7 \times 10^{3}$~K, though there is an offset in the normalization. This offset is due to the presence of the radiative layer. A fully convective model at these high photospheric temperatures would lie on the analytic track:  an example of such a model calculated for T$_{\rm ph} =9000$ is plotted in Fig.~\ref{fig:m_vs_tph} as a star. For higher $m_{\rm BH}$, the offset is reduced as the radiative layer gets thinner.  At lower temperatures, $m_*$ deviates slowly from the power-law until $T_{\rm min}$ is approached, whereupon $m_*$ drops more steeply.  This is the catastrophic shrinking of the quasistar prior to evaporation.  In this regime the analytic scalings do not hold.

\begin{figure}
\psfig{file=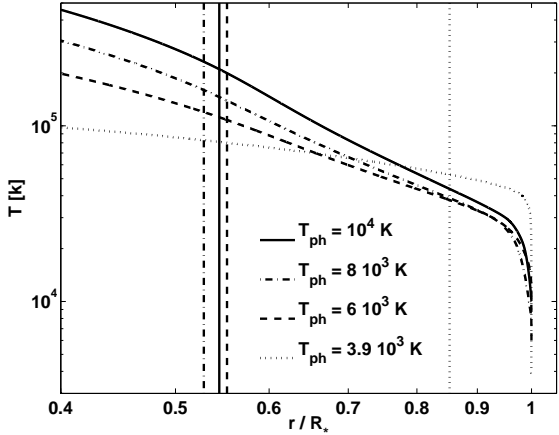,width=0.48\textwidth}
\caption[]{Radial profiles of the interior temperature for the four models plotted in  Fig.~\ref{fig:kappat}. For each model the transition 
radius $R_{\rm tr}$ is marked by the corresponding vertical line.}
\label{fig:tr}
\end{figure}
\begin{figure}
\psfig{file=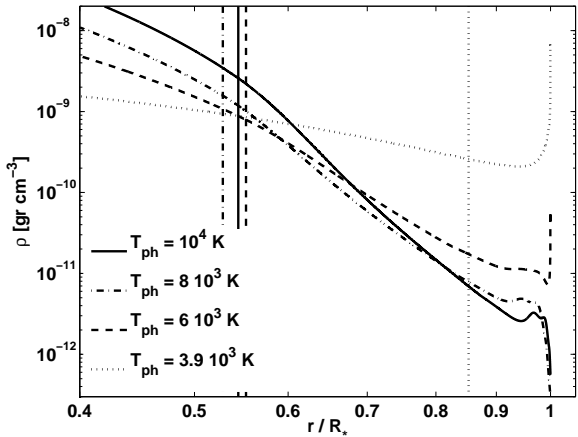,width=0.48\textwidth}
\caption[]{The same as Fig.~\ref{fig:tr}, but for density. As $T_{\rm ph}$ decreases, the radiative layer becomes denser.}
\label{fig:rhor}
\end{figure}

Although the {\em integrated} properties of the numerical Pop~III opacity models resemble those derived analytically and numerically using the toy opacity, the presence of locally super-Eddington fluxes for photospheric temperatures $T_{\rm ph} > T_{\rm min}$ has substantial implications for the derived structures. Fig.~\ref{fig:tr} and Fig.~\ref{fig:rhor} show the radial profiles of temperature and density for the models shown in Fig.~\ref {fig:kappat} and Fig.~\ref{fig:lradleddt}. In hydrostatic equilibrium, the combination of equations~(\ref{eq:hye}) and (\ref{eq:eqstate}) yields an expression for the density gradient, 
\begin{equation}
{kT\over \mu} \frac{d\rho}{dr} =  - \frac{G M(r)}{r^2} \rho - \frac{dp_r}{dr} -  \frac{\rho k}{\mu} \frac{dT}{dr}.
\label{eq:drhodr}
\end{equation}
In regions where the radiative force $-{dp_r}/{dr}$ substantially exceeds the gravitational force the density gradient becomes positive and a local density inversion forms. At the same time, the temperature gradient steepens. In turn, the increase in the density and the decrease of temperature enhance the contribution from bound-free hydrogen absorption, raising the opacity further above $\kappa_{\rm es}$. This general behavior becomes more 
marked as the photospheric temperature drops, and the radiative layer becomes denser and cooler.

The existence of density inversions in our models immediately raises the question of whether such structures are stable. There are two conceptually distinct concerns. In {\em one dimension}, the hydrostatic configurations we have computed are self-consistent, provided that the super-Eddington regions lie beneath the photosphere. An inward-directed force due to the gas pressure gradient compensates for the imbalance between radiation pressure and gravity. However, there could exist a second solution in which the super-Eddington flux drives a wind from the quasistar. If such solutions 
exist, the hydrostatic solution could be unstable even in one dimension, leading to mass loss at temperatures well above $T_{\rm min}$. In {\em two dimensions}, it is even more doubtful that a density inversion would be stable. The resulting instabilities could lead to lateral density contrasts and non-magnetized photon bubbles of the type analyzed by \cite{shaviv01}.

We have not had notable success in analytically estimating the mass loss rate that might occur due to locally super-Eddington fluxes in the quasistar, and our numerical scheme is not suited to tackle the problem. This failure is hardly surprising, since the problem of what happens when a star develops a limited super-Eddington zone has been studied extensively in the context of Luminous Blue Variables such as $\eta$~Carina (for a review, see Humphreys \& Davidson 1994), apparently without any definitive theoretical resolution being attained. One possibility is that a large-scale circulation pattern
develops, superficially resembling convection, and the star suffers no mass loss at all \citep{owocki04}. Another possibility, empirically favored for LBVs, is that episodic mass loss occurs. 

For what follows, we conjecture that since the super-Eddington zones in our models arise due to high densities (and vanish if the density is reduced), quasistars with Pop~III opacities may develop outer regions in which there is circulation but little or no mass loss. As the width of the super-Eddington region becomes comparable to the radiative layer thickness, however, the extended acceleration zone is likely to permit a strong wind to develop. We therefore adopt $T_{\rm min}$ as an estimate for the minimum temperature a quasistar can sustain before evaporating. That there are uncertainties in this identification should be very obvious.


\section{Co-Evolution of Black Holes and Quasistars}
\label{sec:evolution}
The properties of a quasistar will change as the black hole grows and, possibly, as the quasistar itself accretes matter from its environment.  The thermal timescale in a quasistar interior is sufficiently short that the structure can adjust quasistatically. Therefore, we can model the co-evolution of the black hole and envelope as a series of equilibrium models, as long as the photospheric temperature exceeds the floor associated with the opacity
crisis, $T_{\rm ph} > T_{\rm min}$.

We will consider two models for the quasistar's interaction with its environment. In the first, we will assume that the quasistar accretes matter at a constant rate, which we parameterize as $0.1 \dot m_{0.1} M_\odot$ yr$^{-1}$.  This growth rate is consistent with the BVR scenario for black hole growth in pregalactic halos. In the second, we will assume that the quasistar has a fixed mass, which we parameterize in units of $10^6 M_\odot$. We normalize time in years and assume, again 
motivated by the BVR arguments, that the initial seed black hole mass is much smaller than the final mass attained 
prior to dissolution of the quasistar.

\subsection{Before the Opacity Crisis}
As long as $T_{\rm ph} > T_{\rm min}$, the growth rate of the black hole is set by the Eddington limit of the quasistar, implying
\begin{equation}
\label{mdote}
\dot m_{\rm BH} = 2.5 \times 10^{-8} \epsilon_{0.1}^{-1} m_* M_\odot \ {\rm yr}^{-1} , 
\end{equation}
where $\epsilon = 0.1 \epsilon_{0.1}$ is the accretion efficiency and we have taken $\ell_{\rm tr} = \tilde\kappa = 1$. The accretion efficiency relates the luminous output due to accretion with the rate of growth of the black hole mass. This efficiency factor is distinct from the Bondi efficiency factor $\alpha$ defined in eq.~(\ref{eq:lbh}), and could be smaller than the standard value of 0.1 if energy is lost mechanically in a jet that does not couple to the envelope. When the quasistar grows at a steady rate, we have $m_* = 0.1 \dot m_{0.1} t_{\rm yr}$ and the black hole mass grows according to 
\begin{equation}
m_{\rm BH} = 1.2 \times 10^{-9} \epsilon_{0.1}^{-1}  \dot m_{0.1} t_{\rm yr}^2  = 1.2 \times 10^{-7} \epsilon_{0.1}^{-1}\dot m_{0.1}^{-1} m_*^2  . 
\label{eq:mbh1}
\end{equation}
If the quasistar has a fixed mass, then the black hole grows linearly with time:
\begin{equation}
\label{mbh2}
m_{\rm BH} = 2.5 \times 10^{-2} \epsilon_{0.1}^{-1}  m_{*6} t_{\rm yr}. 
\end{equation} 
Equations (\ref{eq:t_m_bh}) and (\ref{eq:mbh1}) imply that the photospheric temperature decreases as the black hole grows. The quasistar thus evolves toward the opacity crisis, reaching it when $T_{\rm ph} = T_{\rm min} \equiv 4000 T_{\rm m,4}$ K. For the case of steady quasistar growth, this occurs when the quasistar mass reaches 
\begin{equation}
\label{mbh3}
m_{\rm *0} = 1.8 \times 10^{5} \epsilon_{0.1}^{8/9} \dot m_{0.1}^{-8/9} \alpha_{0.1}^{-4/9}  T_{m,4}^{-20/9} 
\end{equation}
and the black hole mass reaches
\begin{equation}
\label{mbh4}
m_{\rm BH,0} = 3.9\times 10^{3} \epsilon_{0.1}^{7/9} \dot m_{0.1}^{-7/9} \alpha_{0.1}^{-8/9}  T_{m,4}^{-40/9} .
\end{equation}
For a fixed envelope mass, the opacity crisis is reached when 
\begin{equation}
\label{mbh5}
m_{\rm BH,0} = 1.9 \times 10^{4} \alpha_{0.1}^{-1/2} m_{*6}^{7/8} T_{m,4}^{-5/2} . 
\end{equation}


\subsection{Numerical Results for Quasistar Evolution}

\begin{figure}
\psfig{file=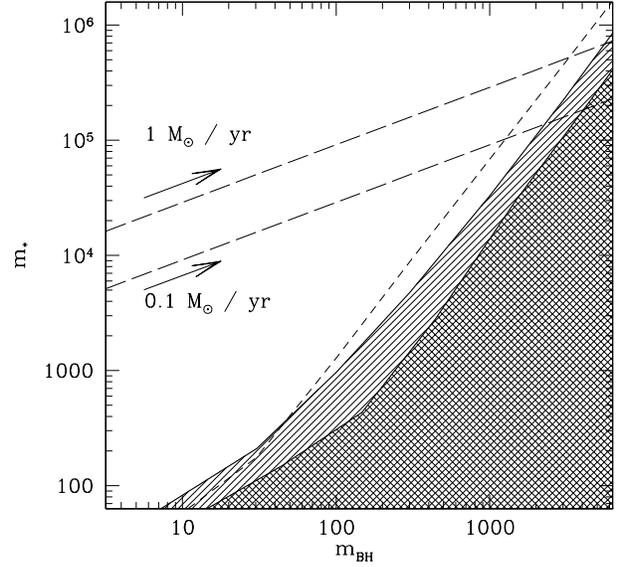,width=0.48\textwidth}
\caption[]{Envelope mass versus black hole mass for static solutions at the minimum photospheric temperature. The dashed line is for the ``toy" opacity model, the solid lines are for the numerical opacity. Static solutions are excluded in the lower shaded regions. The lighter shaded region is computed assuming that $\alpha =0.1$, the darker shaded region is for $\alpha =0.05$.  Superimposed on the figure are the evolutionary tracks (eq.~[\ref{eq:mbh1}]). The upper track is for an accretion rate onto the envelope of $\dot{M}_* $= 1 $\ M_{\odot} \ {\rm yr}^{-1}$.  The lower track is for $\dot{M}_*$= 0.1$ \ M_{\odot}\ {\rm yr}^{-1}$.}
\label{fig:ms_vs_mbh}
\end{figure}

To test the above results, we numerically computed explicit evolutionary sequences by solving the 
equations,
\begin{eqnarray} 
 \frac{{\rm d}M_{\rm BH}}{{\rm d}t} &=& \frac{L_{\rm BH}}{\epsilon c^2} \\
 \frac{{\rm d}M_*}{{\rm d}t} &=& \dot{M}_*,
\end{eqnarray}  
with $L_{\rm BH} (M_*, M_{\rm BH}, \alpha)$ calculated from the model with no assumption as to the luminosity relative to the Eddington limit. The analytic evolution tracks for quasistars depend only on the assumption that the luminosity of the quasistar is close to the Eddington limit appropriate to the total mass.  Our numerical results verified that this is a very good approximation. Accordingly, to estimate the maximum mass a black hole can grow to within a quasistar it suffices to use 
the analytic growth track given by equation~(\ref{eq:mbh1}). The analytic estimate of the minimum temperature is 
less accurate. We therefore combine the analytic growth track with the numerically computed minimum photospheric temperature to derive the final mass.
 
In Fig.~\ref{fig:ms_vs_mbh}, we plot analytic tracks corresponding to envelope accretion rates of 0.1~$M_\odot \ {\rm yr}^{-1}$ and 
1~$M_\odot \ {\rm yr}^{-1}$. As the black hole grows, the tracks move toward the right-hand side of the plot, eventually crossing into the forbidden region set by the minimum photospheric temperature.  The figure shows that the final black hole mass is higher for higher accretion rates onto the envelope, because the effective temperature decreases more slowly, delaying the final dissolution at $T_{\rm min}$. We find that quasistars are indeed an efficient channel for ``growing'' intermediate mass black holes. For $\alpha = 0.1$ and $\dot m_{0.1} \ge 1$, the final black hole mass is predicted to be at least a few thousand Solar masses. 


\subsection{Scaling of Results with $\alpha$}
\label{sec:alpha}
The parameter $\alpha$, which accounts for inefficiencies in the accretion flow within the Bondi radius, is largely unknown. The numerical results we have calculated are mostly for $\alpha = 0.1$. Analytically, there is a simple scaling with $\alpha$, valid for $m_{*} \gg m_{\rm BH}$. In this limit, the only dependence of the quasistar structure on the black hole mass enters via eq.~(\ref{eq:lbh}) for the luminosity, $L_{\rm BH} \propto M_{\rm BH}^2 \alpha$. Any solution in which the combination ${\cal{M}} = M_{\rm BH} \alpha^{1/2}$ is constant should then have the same $m_*$ and $T_{\rm ph}$, along with the same radial structure well outside the Bondi radius. If $\alpha$ were to be smaller than our assumed value of 0.1, this would allow larger black holes to grow within the same envelopes. 

Numerically, the expectation that the results depend only on $\cal{M}$ is borne out for $m_{\rm BH} > 10^3$, irrespective of the photospheric temperature. For $m_{\rm BH} < 10^3$, it is valid only for photospheric temperatures well in excess of the minimum temperature. For low black hole masses and photospheric temperatures approaching the minimum value, the black hole mass cannot be ignored in the equation of hydrostatic equilibrium, and the resulting structure depends separately on $m_{\rm BH}$ and $\alpha$. In Fig.~\ref{fig:ms_vs_mbh} we plot numerical results for the minimum photospheric temperature computed with $\alpha = 0.1$ and $\alpha = 0.05$. As is evident from the figure, the offset between these curves is not constant. However, for the higher masses that are of interest when determining the maximum black hole mass that can be attained, the $\alpha^{1/2}$ scaling is quite accurate.


\subsection{Post-Opacity Crisis: Dissolution of the Quasistar}
Our analysis of the junction between the radiative layer and convective zone indicates that no static solutions exist for $T_{\rm ph} < T_{\rm min}$.  In contrast to a red giant, protostar, or Thorne-\.Zytkow object, where feedback allows the energy source or envelope to adjust so that the photosphere follows the Hayashi track, no stable feedback appears to exist here.  Once a quasistar reaches $T_{\rm min}$, the plummeting opacity causes the convective zone to release radiation at a super-Eddington rate.  The deflation of the convective zone increases the rate of accretion onto the black hole, leading to a runaway.  The only ways to avoid the destruction of the quasistar would be to decrease the mass of the black hole or to increase the mass of the envelope at an unrealistically high rate.

A detailed analysis of the mass loss process is beyond the scope of this paper, but it is easy to show that the black hole is unlikely to grow much while the quasistar is evaporating.  Once mass loss starts in earnest, the wind will carry away nearly all the energy released by the black hole.  Assuming that the wind speed is of order the escape speed from the quasistar, $v_W \sim (GM_*/R_*)^{1/2}$, this implies that the quasistar evaporates when the energy liberated by accretion equals the binding energy. (Although the quasistar is mainly radiation pressure-dominated, we assume that the binding energy is enhanced by rotation.) The mass accumulated by the black hole during the dissolution phase is then
\begin{equation}
\label{bhdiss}
\Delta M_{\rm BH} \sim {GM_{*0}^2 \over \epsilon R_{*0} c^2} \approx 20 \epsilon_{0.1}^{1/3} \dot m_{0.1}^{-4/3} \alpha_{0.01}^{-7/9}  T_{m,4}^{-4/3} \ M_\odot,
\end{equation}
which is negligibly small compared to $m_{\rm BH,0}$.  Therefore, we may regard $m_{\rm BH,0}$ as the maximum mass attainable by a black hole growing inside a quasistar envelope. 


\section{Discussion and Conclusions}
In this paper, we have studied the structure and evolution of quasistars, rapidly accreting black holes embedded within massive gas envelopes. We find that, for any black hole mass, there is a minimum photospheric temperature below which rapid dissolution of the envelope is inevitable. Both analytic and numerical models, computed using Pop~III opacities, suggest that this minimum temperature is around 4000-5000~K. If quasistars are implicated in the formation of seeds for supermassive black holes in pre-galactic haloes, as suggested by \cite{Be06}, this floor temperature implies that the most luminous quasistars would emit most strongly in the rest-frame near-IR. At typical redshifts of $z \simeq 10$, the observed spectrum would peak at $\lambda \sim 10 \mu {\rm m}$. We defer discussion of the possible cosmological density of such sources, and hence their observability with future facilities such as the {\em James Webb Space Telescope}, to a subsequent paper.

We have also studied the evolution of quasistars in simple scenarios for the mass growth of their envelopes. Generically, as the black hole mass grows the photospheric temperature falls, until eventually the limit imposed by the behavior of the Pop~III opacity is reached.  Unveiling of the black hole appears inevitable long before it succeeds in accreting much of the envelope. Both analytic and numerical estimates suggest that seed black holes with masses between a few $10^3 \ M_\odot$ and $10^4 \ M_\odot$ are plausible outcomes of this scenario. The efficiency of the black hole accretion process that powers quasistars (parameterized here by $\alpha$) is unknown --- if the efficiency is low (due, for example, to polar outflows that couple poorly to the envelope) then larger seed black holes are possible. 

Although our numerical integrations of quasistar structure confirm many aspects of the analytic model, they also reveal complex behavior in the radiative zone immediately beneath the photosphere. Narrow regions in which the flux is locally super-Eddington develop for temperatures significantly in excess of the minimum temperature at which the entire radiative zone becomes super-Eddington. In our hydrostatic models, these zones are characterized by density inversions. Interpreting this structure is tricky, since such density inversions may well be subject to one- or two-dimensional instabilities whose ultimate resolution is unknown. Hydrodynamic studies will be needed to determine the extent of mass loss that may occur at temperatures above the theoretical floor value. 

Finally, we note that here we have focused on black holes embedded within truly primordial gas at high 
redshift. {\em Small} amounts of metal pollution would act to increase the opacity in the radiative zone, 
altering the minimum temperature and possibly increasing the likelihood of mass loss due to the formation 
of dust. Changes to the interior structure would be smaller. In particular, the density at the base of the 
radiative zone is so low ($\rho \sim 10^{-9} \ {\rm g cm}^{-3}$) that electron scattering will continue to dominate the 
opacity for $T > 10^5 \ {\rm K}$, even in the presence of pollution. Of greater concern is the fact  
that metal enriched gas in the halo {\em outside} the quasistar would be more susceptible to fragmentation 
and star formation. Too much star formation would reduce the rate of mass accretion onto the quasistar below 
the values ($\sim 0.1 \ M_\odot {\rm yr}^{-1}$ or higher) that we have assumed, resulting in smaller 
final black hole masses. It is also possible that generically similar structures could form at lower redshift whenever small black holes encounter very high rates of gas inflow. Structures similar to those we have described could allow stellar remnants at the center of merging galaxies to grow significantly via a transient quasistar stage.

\section*{Acknowledgments}
We thank Michael Mayer for providing us with high resolution tables of Pop~III opacities. MCB and PJA acknowledge support from NASA's Astrophysics Theory Program under grants NNG04GL01G and NNX07AH08G; from NASA's Beyond Einstein Foundation Science Program under grant NNG05GI92G; and from the NSF under grants AST~0307502 and AST~0407040. EMR acknowledges support from NASA though Chandra Postdoctoral Fellowship grant number PF5-60040 awarded by the Chandra X-ray Center, which is operated by the Smithsonian Astrophysical Observatory for NASA under contract NASA8-03060.

\end{document}